\documentclass[aps,reprint,groupedaddress,showpacs,
]{revtex4-1}

\usepackage{amssymb,amsmath}
\usepackage{graphicx}
\usepackage{xcolor}
\usepackage{dsfont}

\usepackage[english]{babel}

\usepackage{siunitx}

\makeatletter
\renewcommand{\fnum@figure}{\textbf{Fig. \thefigure}}
\makeatother

\makeatletter
\renewcommand{\fnum@table}{\textbf{Tab. \thetable}}
\makeatother

\newcommand{\Jbold}{\mathbf{J}}

\newcommand{\ket}[1]{\left| #1 \right>} % for Dirac bras
\newcommand{\I}{\mathrm{i}}
\newcommand{\E}{\mathrm{e}}
\newcommand{\cat}{\ket{\psi_{\mathrm{kitten}}}}

\newcommand{\unitb}{\hat{\mathbf{b}}}
\newcommand{\unitu}{\hat{\mathbf{u}}}
\newcommand{\unitv}{\hat{\mathbf{v}}}
\newcommand{\unitx}{\hat{\mathbf{x}}}
\newcommand{\unity}{\hat{\mathbf{y}}}
\newcommand{\unitz}{\hat{\mathbf{z}}}

\newcommand{\muB}{\mu_{\mathrm{B}}}

\usepackage{layouts}
\usepackage{float}

\begin{document}

\title{
Quantum-enhanced sensing using non-classical spin states of a highly magnetic atom
}

\author{Thomas Chalopin}
\author{Chayma Bouazza}
\author{Alexandre Evrard}
\author{Vasiliy Makhalov}
\author{Davide Dreon}
\altaffiliation
{Present address: Department of Physics, ETH Zurich, 8093 Zurich, Switzerland}
\author{Jean Dalibard}
\author{Leonid A. Sidorenkov}
\altaffiliation
{Present address: SYRTE, Observatoire de Paris, PSL University, CNRS, Sorbonne Universit\'e, LNE, 61 avenue de l'Observatoire, 75014 Paris, France}
\email{Corresponding author (email: leonid.sidorenkov@obspm.fr).}

\author{Sylvain Nascimbene}

\affiliation{Laboratoire Kastler Brossel,  Coll\`ege de France, CNRS, ENS-PSL University, Sorbonne Universit\'e, 11 Place Marcelin Berthelot, 75005 Paris, France}

\date{\today}
\begin{abstract}
\noindent 
Coherent superposition states of a mesoscopic quantum object play a major role in our understanding of the quantum to classical boundary, as well as in quantum-enhanced metrology and computing. However, their practical realization and manipulation remains challenging,  requiring a high degree of control of the system and its coupling to the environment. Here, we use dysprosium atoms -- the most magnetic element in its ground state  -- to realize coherent superpositions between electronic spin states of opposite orientation, with a mesoscopic spin size $J=8$. We drive coherent spin states to quantum superpositions using non-linear light-spin interactions, observing a series of collapses and revivals of quantum coherence. These states feature highly non-classical behavior, with a sensitivity to magnetic fields enhanced by a factor 13.9(1.1) compared to coherent spin states -- close to the Heisenberg limit $2J=16$ -- and an intrinsic fragility to environmental noise.

\end{abstract}

\maketitle

\noindent  Future progress in quantum technologies is based on the engineering and manipulation of physical systems with highly non-classical behavior \cite{frowis_macroscopic_2018}, such as quantum coherence \cite{streltsov_colloquium:_2017}, entanglement \cite{horodecki_quantum_2009}, and quantum-enhanced metrological sensitivity \cite{giovannetti_advances_2011,pezze_non-classical_2016}. These properties generally come together with an inherent fragility due to decoherence via the coupling to the environment, which makes the generation of highly non-classical states challenging \cite{zurek_decoherence_2003,schlosshauer_decoherence:_2007}.
An archetype of such systems consists in an object prepared in a coherent superposition of two distinct quasi-classical states, realizing a conceptual instance of Schr\"odinger cat \cite{haroche_exploring_2006}. Such states have been realized in systems of moderate size -- referred to as `mesoscopic' hereafter -- with trapped ions \cite{monroe_schrodinger_1996,monz_14-qubit_2011}, cavity quantum electrodynamics (QED)  systems \cite{brune_observing_1996,deleglise_reconstruction_2008,facon_sensitive_2016}, superconducting
quantum interference devices \cite{friedman_quantum_2000}, optical photons  \cite{ourjoumtsev_generating_2006,neergaard-nielsen_generation_2006,takahashi_generation_2008,yao_observation_2012}   and  circuit QED systems \cite{kirchmair_observation_2013,vlastakis_deterministically_2013}. Non-classical behavior can also be achieved with other types of quantum systems, including squeezed states  \cite{esteve_squeezing_2008,gross_nonlinear_2010,riedel_atom-chip-based_2010,maussang_enhanced_2010,lucke_twin_2011,hamley_spin-nematic_2012,berrada_integrated_2013,lucke_detecting_2014,bohnet_reduced_2014,hosten_measurement_2016,cox_deterministic_2016}.

Inspired by the hypothetical cat state $\ket{\mathrm{dead}}+\ket{\mathrm{alive}}$ introduced by Schr\"odinger in his famous Gedankenexperiment, one usually refers to a cat state
in quantum optics as a superposition of quasi-classical states  consisting in coherent states of the electromagnetic field, well separated in phase space and playing the role of the $\ket{\mathrm{dead}}$ and $\ket{\mathrm{alive}}$ states \cite{haroche_exploring_2006}. Mesoscopic cat states can be dynamically generated in photonic systems, e.g. using a Kerr non-linearity \cite{yurke_generating_1986,kirchmair_observation_2013}. For a spin $J$, a quasi-classical coherent state is represented as a state $\ket{\pm J}_{\unitu}$ of maximal spin projection $m=\pm J$ along an arbitrary direction $\unitu$. Such a state constitutes the best possible realization of a classical state of well defined polarization, as it  features isotropic fluctuations of the perpendicular spin components, of minimal variance $\Delta J_{\unitv}=\sqrt{J/2}$ for $\unitv\perp\unitu$ \cite{radcliffe_properties_1971}. A cat state can then be achieved for large $J$ values, and it consists in the coherent superposition of two coherent spin states of opposite magnetization, which are well separated in phase space. Such states can be created under the action of non-linear spin couplings  \cite{cirac_quantum_1998,molmer_multiparticle_1999,gordon_creating_1999,castin_bose-einstein_2001}. These techniques have been implemented with individual alkali atoms, using laser fields  to provide almost full control over the quantum state of their hyperfine spin \cite{smith_continuous_2004,chaudhury_quantum_2007,fernholz_spin_2008,smith_quantum_2013,schafer_experimental_2014}. However, the small spin size involved in these systems intrinsically limits the achievable degree of non-classical behavior.

Non-classical spin states have  also been created in  ensembles of one-electron atoms \cite{pezze_non-classical_2016}. When each atom carries a spin-1/2 degree of freedom, a set of $N$ atoms can  collectively behave as
  an effective spin $J=N/2$, that can be driven into  non-classical states via the interactions between atoms \cite{kitagawa_squeezed_1993}. In such systems, while spin-squeezed states have been realized experimentally \cite{esteve_squeezing_2008,gross_nonlinear_2010,riedel_atom-chip-based_2010,maussang_enhanced_2010,lucke_twin_2011,hamley_spin-nematic_2012,berrada_integrated_2013,hosten_measurement_2016}, cat states remain out of reach due to their extreme sensitivity to perturbations (e.g. losing a single particle fully destroys their quantum coherence) \cite{lau_proposal_2014}.

\begin{figure}[]
\hspace*{-0.\linewidth}\includegraphics[width=\linewidth]{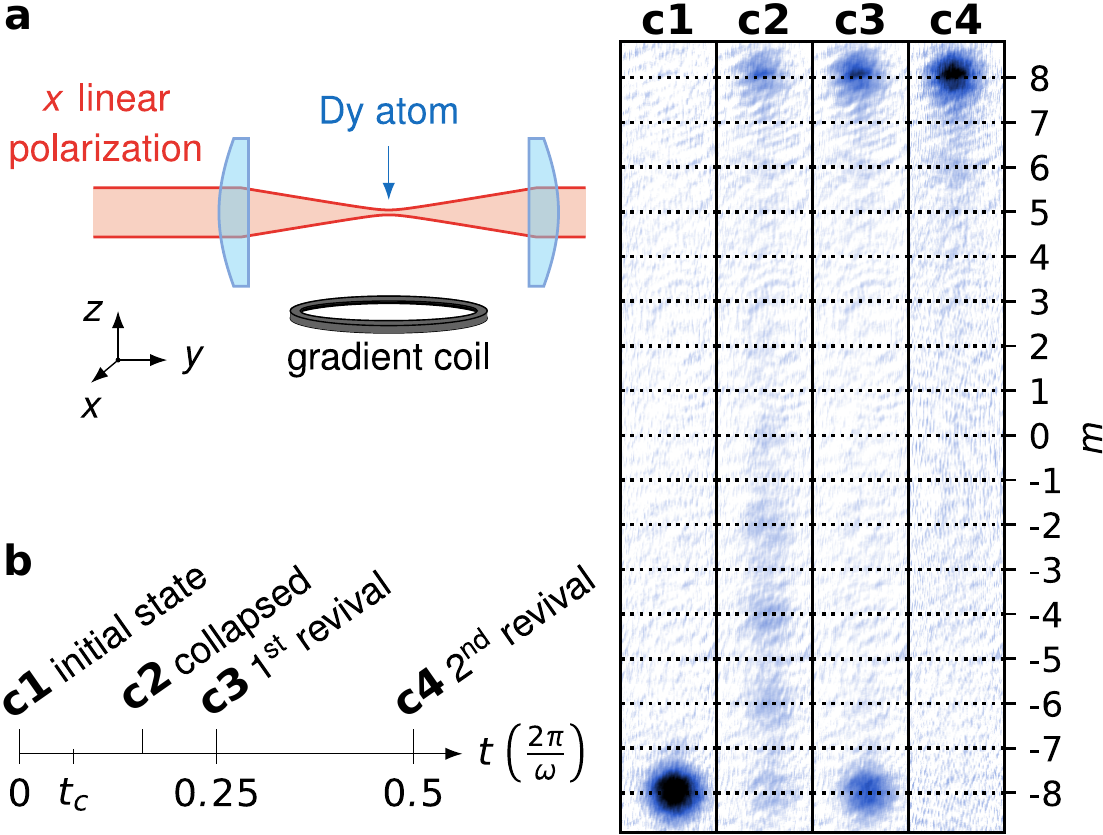}
\vspace{-5mm}
\caption{
\textbf{Experimental scheme and expected spin dynamics.}
(\textbf{a}) Experimental scheme. The spin $J=8$ of Dy atoms is manipulated using an off-resonant laser field linearly polarized along $x$, leading to a non-linear coupling  $\hbar\omega\hat J_x^2$. The spin state is subsequently probed by imaging the atoms after a Stern-Gerlach separation of magnetic sublevels $\ket{m}_z$, allowing to determine their individual populations.
(\textbf{b}) Expected spin dynamics.  The spin, initially prepared in $\ket{-J}_z$ (corresponding atom image (\textbf{c1})), first collapses to a featureless state (\textbf{c2}) on a fast timescale $t_c\ll1/\omega$. It subsequently undergoes a series of revivals, corresponding to the formation of a superposition between states $\ket{-J}_z$ and $\ket{J}_z$  (\textbf{c3}) and later to the polarized state $\ket{J}_z$ (\textbf{c4}). Each image is the average of typically 10 resonant absorption images.
\label{fig_scheme}}
\end{figure}

In this work, we use samples of  dysprosium atoms, each of them carrying an electronic spin of mesoscopic size $J=8$. We exploit the  AC Stark shift produced by off-resonant light \cite{smith_continuous_2004} to drive non-linear spin dynamics (see Fig.~\ref{fig_scheme}). Each atomic spin independently evolves  in a Hilbert space of dimension $2J+1=17$, much smaller than the dimension $2^N\sim10^5$ of an equivalent system of $N=16$ spins 1/2. We achieve the production of quantum superpositions of effective size 13.9(1.1) (as defined hereafter), close to the maximum allowed value $2J=16$ for a spin $J$. 
 We  provide a tomographic reconstruction of the full density matrix of these states and monitor their decoherence due to the dephasing induced by magnetic field noise.

\begin{figure*}
\hspace*{-0.01\linewidth}\includegraphics[width=1.02\linewidth]{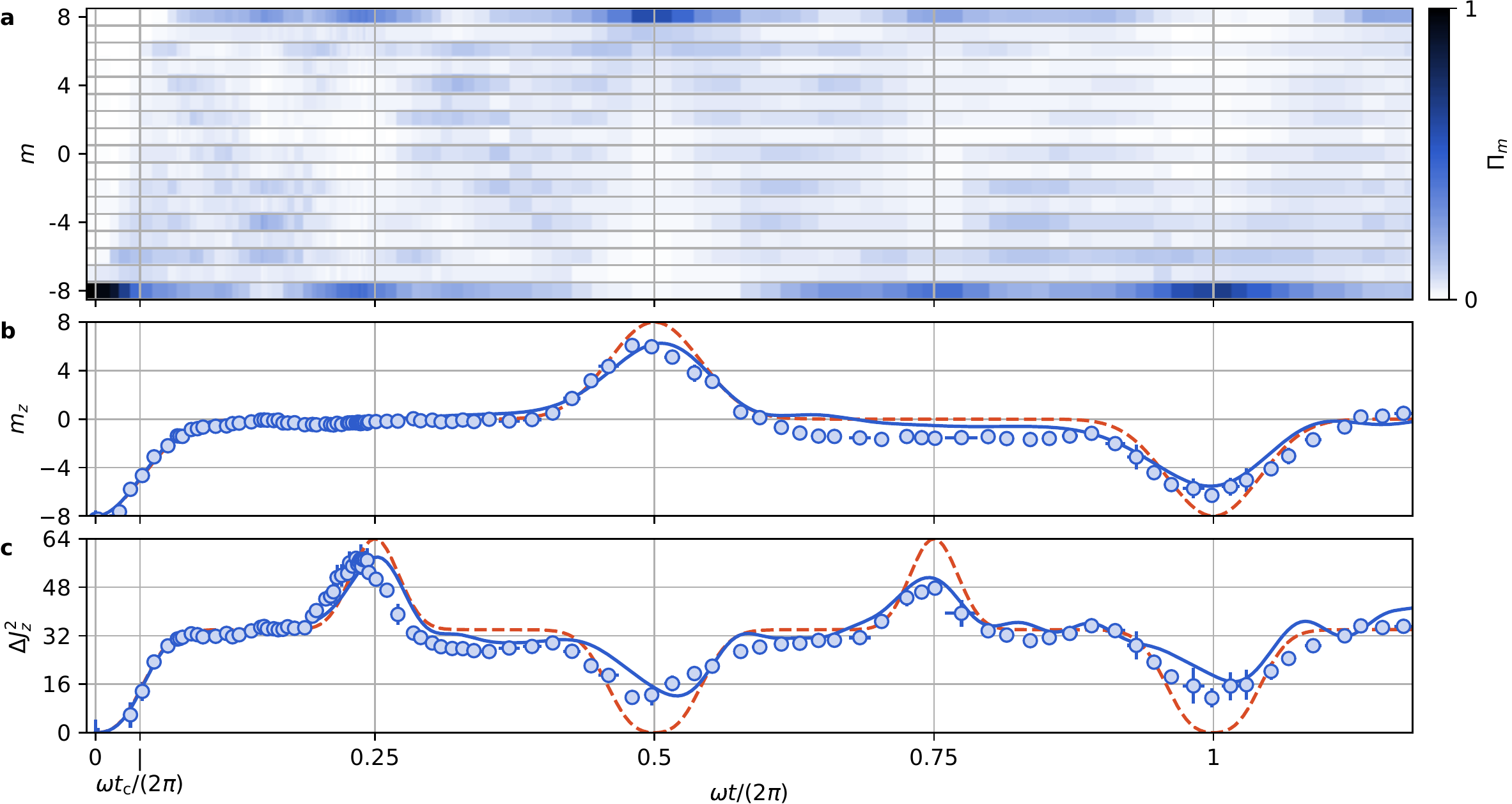}
\vspace{-5mm}
\caption{
\textbf{Collapses and revivals in the non-linear spin dynamics.}
(\textbf{a}) Evolution of the spin projection probabilities $\Pi_m$ along $z$ as a function of the light pulse area $\omega t$.  
(\textbf{b}, \textbf{c}) Evolution of the magnetization $m_z$  and of the variance in the spin projection $\Delta J_z^2$ calculated from the $\Pi_m$ distributions.
 The dashed red lines represent the ideal evolution for a $\hat J_x^2$ coupling \cite{kitagawa_squeezed_1993}, and the blue solid lines correspond to a fit taking into account experimental imperfections (see Methods). Each point is the average of 5 measurements, and the error bars represent the 1$\sigma$ statistical uncertainty.
\label{fig_evolution}}
\end{figure*}

\bigskip
\noindent\textbf{RESULTS}

\noindent Our experimental scheme is sketched in Fig.~\ref{fig_scheme}a. We use an ultracold sample of about $10^5$ $^{164}$Dy atoms, initially spin-polarized in the absolute ground state $\ket{-J}_z$, under a quantization magnetic field $\mathbf{B}=B\unitz$, with \mbox{$B=18.5(3)\,$mG} (see Methods).  The non-linear spin dynamics is induced by a laser beam focused on the atomic sample. The laser wavelength is chosen close to the $626$-nm resonance line so as to produce light shifts with a strong spin dependence. For a linear light polarization along $x$, we expect the spin dynamics to be described by the Hamiltonian \cite{smith_continuous_2004}
\begin{equation}\label{eq_H}
\hat H=\hbar\omega_{\mathrm{L}}\hat J_z+\hbar\omega \hat J_x^2,
\end{equation}
where the first term corresponds to the Larmor precession induced by the magnetic field, and the second term is the light-induced spin coupling. The light beam intensity and detuning from resonance are set such that the light-induced coupling frequency \mbox{$\omega=2\pi\times1.98(1)\,$MHz} largely exceeds the Larmor precession frequency \mbox{$\omega_{\mathrm{L}}=2\pi\times31.7(5)$\,kHz}. In such a regime the Hamiltonian of Eq.~\ref{eq_H} takes the form of the so-called one-axis twisting Hamiltonian, originally introduced for generating spin squeezing \cite{kitagawa_squeezed_1993,gross_nonlinear_2010,riedel_atom-chip-based_2010}. We drive the spin dynamics using light pulses of duration $t\sim10\,$ns to \SI{1}{\micro\second}. Once all laser fields are switched off, we perform a projective measurement of the spin along the $z$ axis in a Stern-Gerlach experiment (see Fig.~\ref{fig_scheme}c).  Measuring the atom number corresponding to each projection value $m$ allows to infer the projection probabilities $\Pi_m$, $-J\leq m\leq J$. 

\medskip
%\filbreak
\noindent\textbf{Quantum state collapses and revivals}

\noindent We first investigated the evolution of the spin projection probabilities $\Pi_m$ as a function of the light pulse duration $t$. As shown in Fig.~\ref{fig_evolution}, we find the spin dynamics  to involve mostly the even $\ket{m}_z$ states. This behavior is expected from the structure of the  $\hat J_x^2$ coupling, which does not mix the even- and odd-$\ket{m}_z$ sectors. 

Starting in $\ket{-J}_z$, we observe for short times that all even-$\ket{m}_z$ states get gradually populated. The magnetization $m_z\equiv\langle\hat J_z\rangle$ and spin projection variance $\Delta J_z^2$ relax to almost constant values  $m_z=-0.3(2)$ and $\Delta J_z^2=33(1)$ in the whole range  $0.2\,\pi<\omega t<0.36\,\pi$. This behavior agrees with the expected collapse of  coherence induced by a  non-linear coupling. To understand its mechanism in our system, we write the initial state in the $x$ basis, as 
\begin{equation}\label{eq_i}
\!\ket{-J}_z=\sum_m(-1)^mc_m\ket{m}_x, \quad \!c_m=2^{-J}\sqrt{{{2J}\choose{J\!+\!m}}}.
\end{equation}
In this basis,  the non-linear coupling $\hat J_x^2$ induces $m$-dependent phase factors, leading to the state
\begin{equation}\label{eq_psi}
\ket{\psi(t)}=\sum_m(-1)^m\E^{-\I m^2\omega t}c_m\ket{m}_x.
\end{equation}
The variations between the accumulated phase factors lead to an apparent collapse of the state coherence \cite{cummings_stimulated_1965}. The collapse timescale $t_{\mathrm{c}}$ can be estimated by calculating the typical relaxation time of the magnetization, yielding $t_{\mathrm{c}}=1/(\sqrt{2J}\omega)$, i.e. $\omega t_{\mathrm{c}}=0.08\,\pi$ \cite{kitagawa_squeezed_1993,castin_bose-einstein_2001} (see the Supplementary Materials). 

For longer evolution times, we observe the occurence of peaks in $m_z(t)$ or $\Delta J_z^2$, that we interpret as quantum coherence revivals \cite{eberly_periodic_1980,rempe_observation_1987,kirchmair_observation_2013}. After a quarter of the period, i.e. $\omega t = \pi/2$, all odd-$m$ (and all even-$m$) phase factors in Eq.~\ref{eq_psi} get in phase again, leading to the superposition
\begin{equation}\label{eq_cat}
\cat=\E^{\I\pi/4}(\ket{-J}_z - \I \ket{J}_z)/\sqrt{2},
\end{equation}
between maximally polarized states of opposite orientation \cite{molmer_multiparticle_1999,castin_bose-einstein_2001}. Considering the spin $J=8$ as a mesoscopic quantum system, we refer to this state as a Schr\"odinger `kitten' state \cite{ourjoumtsev_generating_2006}. We observe that, for durations  \mbox{$0.45\,\pi<\omega t<0.49\,\pi$}, the magnetization remains close to zero while the variance in the spin projection features a peak of maximal value \mbox{$\Delta J_z^2=57.1(2)$} (see Fig.~\ref{fig_evolution}). 

For pure quantum states, such a large variance is characteristic of  coherent superpositions between states of very different magnetization. However, from this sole measurement we cannot exclude the creation of an incoherent mixture of $\ket{\pm J}_z$ states.
We observe at later times additional revivals of coherence that provide a first evidence that the state discussed above indeed corresponds to a coherent quantum superposition. The second revival occurs around $\omega t=\pi$, and corresponds to a re-polarization of the spin up to $m_z=6.0(1)$, with most of the atoms occupying the state $\ket{J}_z$. We detect additional revivals around $\omega t=3\pi/2$ and $\omega t = 2\pi$,  respectively corresponding  to another superposition state (large spin projection variance $\Delta J_z^2=47.0(6)$) and to a magnetized state close to the initial state ($m_z=-6.0(2)$).

The observed spin dynamics qualitatively agrees with the one expected for a pure $\hat J_x^2$ coupling \cite{kitagawa_squeezed_1993}, while a more precise modelling of the data -- taking into account the linear Zeeman coupling produced by the applied magnetic field, as well as experimental imperfections (see Methods) --  matches well our data (see Fig.~\ref{fig_evolution}).

\medskip
\noindent\textbf{Probing the coherence of the superposition}

\begin{figure}
\hspace*{-0.0\linewidth}
\includegraphics[width=1.0\linewidth]{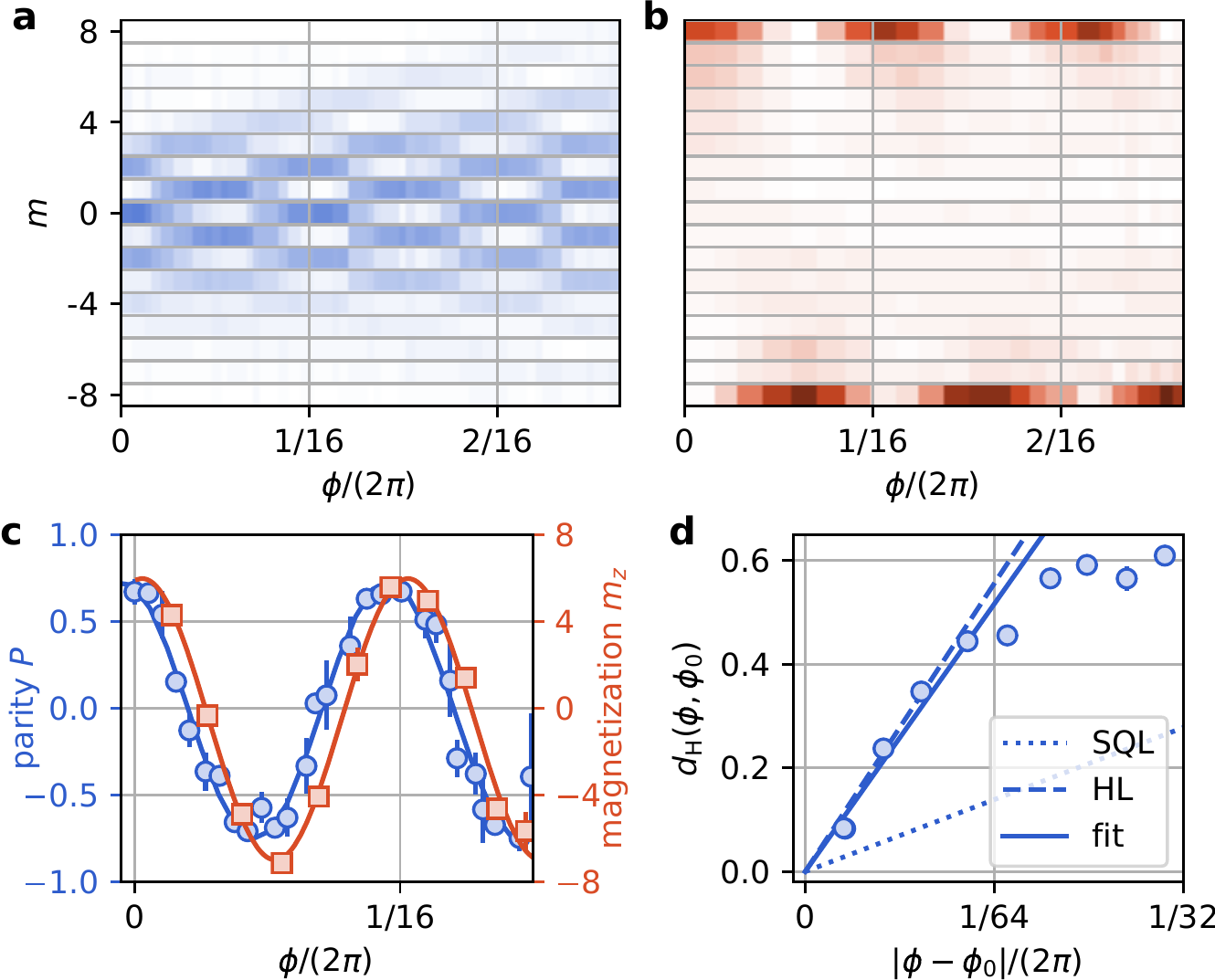}
\vspace{-5mm}
\caption{\textbf{Probing quantum coherence.}
(\textbf{a}) Measured spin projection probabilities $\Pi_m(\phi)$ on the equatorial directions of azimutal angle $\phi$.
(\textbf{b}) Measured spin projection probabilities $\Pi_m(\phi)$ along $z$, after a Larmor rotation of angle $\phi$ followed by a  light pulse inducing further non-linear spin dynamics.
(\textbf{c}) Evolution of the mean parity $P$ (blue circles) and magnetization $m_z$ (red squares) calculated from the probabilities shown in a and b, respectively. The phase shift between the measured oscillations stems from the phase offset associated with the  Larmor rotation around $z$ occurring during the parity measurement sequence.
(\textbf{d}) Variation of the  Hellinger distance between projection probabilities of angles $\phi$ and $\phi_0$ as a function of the relative angle $\phi-\phi_0$, calculated from the data plotted in a  with $\phi_0=0.02$. The solid line corresponds to the linear variation for small angle differences fitted to the data. The dotted (dashed) line corresponds to the  standard quantum limit (Heisenberg limit). Error bars represent the 1$\sigma$ statistical error.
\label{fig_Ramsey}}
\end{figure}

\noindent 
In order to directly probe the coherences we follow another experimental protocol allowing us to retrieve the spin projection along directions lying in the $xy$ equatorial plane, corresponding to observables $\hat J_\phi\equiv\cos\phi\,\hat J_x+\sin\phi\,\hat J_y$ (see Methods). 
The coherence of the state $\cat$, involving the opposite coherent states $\ket{\pm J}_z$, cannot be probed using a linear spin observable such as the magnetization, but requires interpreting the detailed structure of the probability distributions $\Pi_m(\phi)$ \cite{bollinger_optimal_1996}. By expanding the coherent states $\ket{\pm J}_z$ on the eigen-basis $\ket{m}_{\phi}$ of the spin component $\hat J_\phi$, we rewrite the  state  as
\begin{equation}\label{cat_equator}
\!\cat=\frac{\E^{\I\pi/4}}{\sqrt{2}}\sum_m\left[\E^{-\I (J\phi+m\pi)}\!-\E^{\I (J\phi+\frac{\pi}{2})}\right]\!c_m\!\ket{m}_{\phi},\hspace*{-2mm}
\end{equation}
where the $c_m$ coefficients were introduced in Eq.~\ref{eq_i}.
For the particular angles $\phi=(p + 1/4)\pi/J$ ($p$ integer), the  two terms in brackets cancel each other for odd $m$ values. Alternatively, for angles $\phi=(p - 1/4)\pi/J$ we expect destructive interferences  for even $m$ \cite{bollinger_optimal_1996,monroe_schrodinger_1996}.  This  behavior can be revealed in the parity of the spin projection 
\begin{equation}\label{eq_parity}
P(\phi)\equiv\sum_m(-1)^m\Pi_m(\phi)=\sin(2J\phi),
\end{equation}
which oscillates with a period $2\pi/(2J)$.

As shown in Fig.~\ref{fig_Ramsey}a, the experimental probability distributions $\Pi_m(\phi)$ feature strong variations with respect to the angle $\phi$. The center of mass of these distributions remains close to zero, consistent with the zero magnetization of the state $\cat$.  We furthermore observe high-contrast parity oscillations agreeing with the above discussion and supporting quantum coherence between the $\ket{\pm J}_z$ components (see Fig.~\ref{fig_Ramsey}c). 

 \footnotetext[5]{The variation of the center of mass of the $\Pi_m(\phi)$ distributions is consistent with a residual magnetization in the prepared kitten state $\mathbf{m}=[0.1(2),0.84(4),-0.2(2)]_{x,y,z}$. }

Information on maximal-order coherences can be unveiled using another measurement protocol, which consists in applying an additional light pulse identical to the one used for the kitten state generation \cite{leibfried_toward_2004}.  When performed right after the first pulse, the second pulse  brings the state $\cat$ to the polarized state $\ket{J}_z$, which corresponds to the second revival occuring around $\omega t=\pi$ in Fig.~\ref{fig_evolution}. An additional wait time between the two pulses allows for a Larmor precession of angle $\phi$ around $z$, leading to the expected evolution 
\begin{align}
\ket{\psi(\phi)}&=\cos(J\phi)\ket{J}_z+\sin(J\phi)\ket{-J}_z,\label{eq_R1}\\
m_z(\phi)&=J\,\cos(2J\phi).\label{eq_R2}
\end{align}
We vary the wait time and measure corresponding probability distributions $\Pi_m(\phi)$ (Fig.~\ref{fig_Ramsey}b) and magnetization $m_z(\phi)$ (Fig.~\ref{fig_Ramsey}c) consistent with Eqs.~\ref{eq_R1} and \ref{eq_R2}, respectively. This non-linear detection scheme reduces the sensitivity to external perturbations, as it transfers information from high-order quantum coherences  onto the magnetization, much less prone to decoherence.

\medskip
\noindent\textbf{A highly sensitive one-atom magnetic probe}

\noindent 
The Larmor precession of the atomic spins in small samples of atoms can be used for magnetometry combining high spatial resolution and high sensitivity \cite{wildermuth_boseeinstein_2005}.
While previous developments of atomic magnetometers were based on alkali atoms, multi-electron lanthanides such as erbium or dysprosium intrinsically provide an increased sensitivity due to their larger magnetic moment, and potentially a substantial quantum enhancement when probing with non-classical spin states \cite{degen_quantum_2017}.

We interpret below the oscillation of the parity $P(\phi)$ discussed in the previous section as the footing of a magnetometer with quantum-enhanced precision, based on  the non-classical character of the kitten state.  According to generic parameter estimation theory, the Larmor phase $\phi$ can be estimated by measuring an observable $\hat{\mathcal{O}}$ with an uncertainty
\begin{equation}\label{eq_Delta_phi}
\Delta\phi=\frac{\Delta \hat{\mathcal O}}{\mathrm{d}\langle\hat{\mathcal{O}}\rangle/\mathrm{d}\phi}
\end{equation}for a single measurement \cite{pezze_entanglement_2009}. Measuring the angle $\phi$ using coherent spin states (e.g. in a Ramsey experiment) leads to a minimum phase uncertainty
$
\Delta\phi_{\mathrm{SQL}}={1}/{\sqrt{2J}},
$
corresponding to the standard quantum limit (SQL).  For an uncertainty limit on phase measurement $\Delta\phi$ we define the metrological gain compared to the SQL  as the ratio $G\equiv(\Delta\phi_{\mathrm{SQL}}/\Delta\phi)^2$, also commonly referred to as the quantum enhancement of  measurement precision \cite{pezze_non-classical_2016}. 
In this framework, the parity oscillation $P(\phi)$ expected from Eq.~\ref{eq_parity} for the state $\cat$ yields a metrological gain $G=2J$, corresponding to the best precision limit \mbox{$\Delta\phi=1/(2J)$} achievable for a spin $J$ -- the Heisenberg limit. From the finite contrast $C=0.74(2)$ of a sine fit of the measured parity oscillation, we deduce a metrological gain $G=2JC^2=8.8(4)$.  

A further increase of  sensitivity can be achieved using the full information given by the measured probability distributions $\Pi_m(\phi)$ (see Fig.~\ref{fig_Ramsey}a and b) \cite{strobel_fisher_2014}. The phase sensitivity is obtained from the rate of change of the probability distribution $\Pi_m(\phi)$ upon a variation of $\phi$, that we quantify using the Hellinger distance 
\mbox{$
d_{\mathrm{H}}^2(\phi,\phi')\equiv\frac{1}{2}\sum_m[\sqrt{\Pi_m(\phi)}-\sqrt{\Pi_m(\phi')}]^2
$}
between the distributions  $\Pi_m(\phi)$ and  $\Pi_m(\phi')$.
For a standard Ramsey experiment using coherent spin states, one expects the scaling behavior $d_{\mathrm{H}}(\phi,\phi')\simeq \sqrt{J/4}|\phi-\phi'|$ for small angle differences. For  the kitten state given by Eq.~\ref{cat_equator}, we expect an increase in the slope of the Hellinger distance by a factor $\sqrt{2J}$, whose square corresponds to the expected metrological gain $G=2J$ at the Heisenberg limit. We show in Fig.~\ref{fig_Ramsey}d the Hellinger distance computed  from the measured distributions $\Pi_m(\phi)$. Its variation for small angle differences yields a metrological gain 
 $G=13.9(1.1)$. We thus find that using the full information from the probability distributions -- rather than using its parity $P(\phi)$ only --  increases the phase  sensitivity.

For a given quantum state used to measure the Larmor phase, we expect the metrological gain to remain bounded by the value of its spin projection variance, as \mbox{$G\leq2\Delta J_z^2/J=14.3(1)$} \cite{pezze_entanglement_2009}. As the measured gain coincides with this bound within error bars, we conclude that the phase measurement based on the Hellinger distance is optimum. We also performed a similar Hellinger distance analysis based on the distributions $\Pi_m(\phi)$ shown in Fig.~\ref{fig_Ramsey}b leading to a comparable metrological gain $G=14.0(9)$. Further increase of sensitivity would require improving the state preparation.

\medskip
%\filbreak
\noindent\textbf{Tomography of the superposition state}

\begin{figure}
\includegraphics[width=1.0\linewidth]{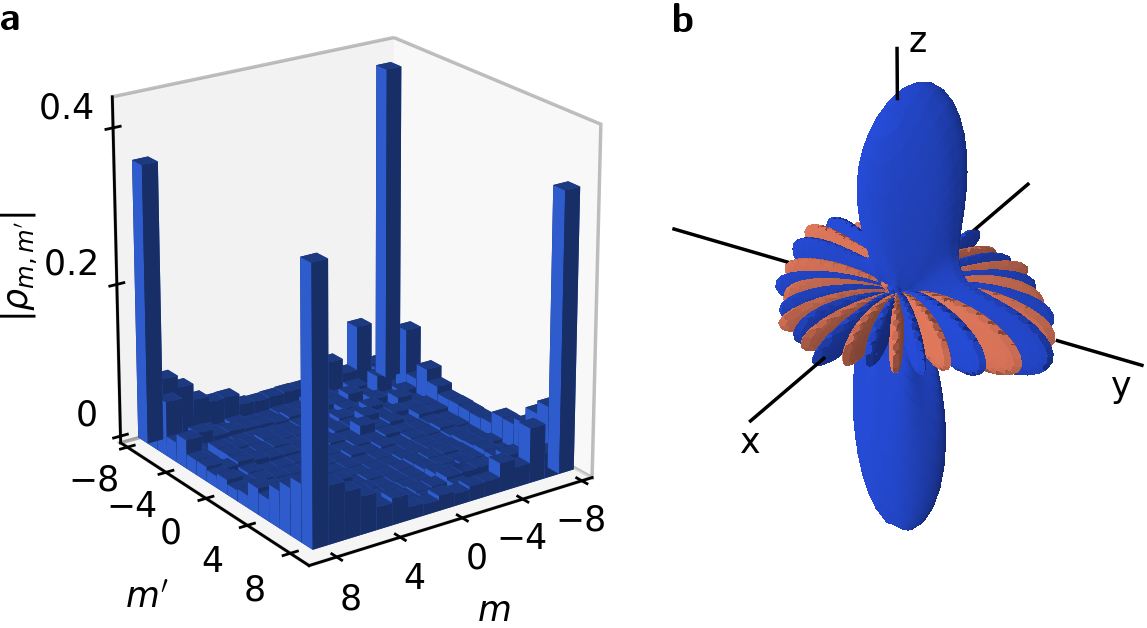}
\vspace{-5mm}
\caption{
\textbf{Tomographic reconstruction of the kitten state.}
(\textbf{a}) Absolute values of the density matrix coefficients $|\rho_{m,m'}|$ fitted from spin projection measurements performed along $z$ and in the equatorial plane. 
(\textbf{b}) Angular Wigner function corresponding to the density matrix plotted in a. Negative-valued regions are plotted in red. 
\label{fig_tomography}}
\end{figure}

\noindent In order to completely characterize the superposition state, we perform a tomographic reconstruction of its density matrix \cite{lvovsky_continuous-variable_2009}. The latter involves \mbox{$(2J+1)^2-1=288$} independent real coefficients, that we determine from a fit of the spin projection probabilities $\Pi_m$ measured on the $z$ axis and on a set of directions uniformly sampling the $xy$ equatorial plane \cite{klose_measuring_2001}.  The inferred density matrix  is plotted in Fig.~\ref{fig_tomography}a. Its strongest elements correspond to populations and coherences involving the coherent states $\ket{\pm J}_z$, as expected for the state $\cat$. We measure  a coherence to population ratio \mbox{$2|\rho_{-J,J}|/(\rho_{-J,-J}+\rho_{J,J})=0.92(8)$}.

In order to further illustrate the non-classical character of the superposition state, we compute from the density matrix its associated Wigner function $W(\theta, \phi)$ \cite{riedel_atom-chip-based_2010},
defined for a spin over the spherical angles $ \theta$, $\phi$  as
\[
W(\theta,\phi)=\sum_{\ell=0}^{2J}\sum_{m=-\ell}^\ell\rho_{\ell}^mY_\ell^m(\theta,\phi),
\]
where $\rho_{\ell}^m$ is the density matrix component on the spherical harmonics $Y_\ell^m(\theta,\phi)$ \cite{dowling_wigner_1994}.
 The reconstructed Wigner function, plotted in Fig.~\ref{fig_tomography}b, exhibits two lobes of positive value around the south and north poles, associated with the population of the states $\ket{\pm J}_z$. It also features interferences around the equatorial plane originating from coherences between these two states, with strongly negative values in a large phase space area. This behavior directly illustrates the highly non-classical character of the kitten state \cite{park_quantum_2016}.

\medskip
\filbreak
\noindent\textbf{Dephasing due to classical noise}

\noindent We furthermore  investigated the environment-induced decay of
quantum coherence by following the evolution of density matrices $\rho(t)$ reconstructed after variable wait times $t$ in the 10-\SI{100}{\micro\second} range. 

\begin{figure}
\hspace*{-0.0\linewidth}\includegraphics[width=1.0\linewidth]{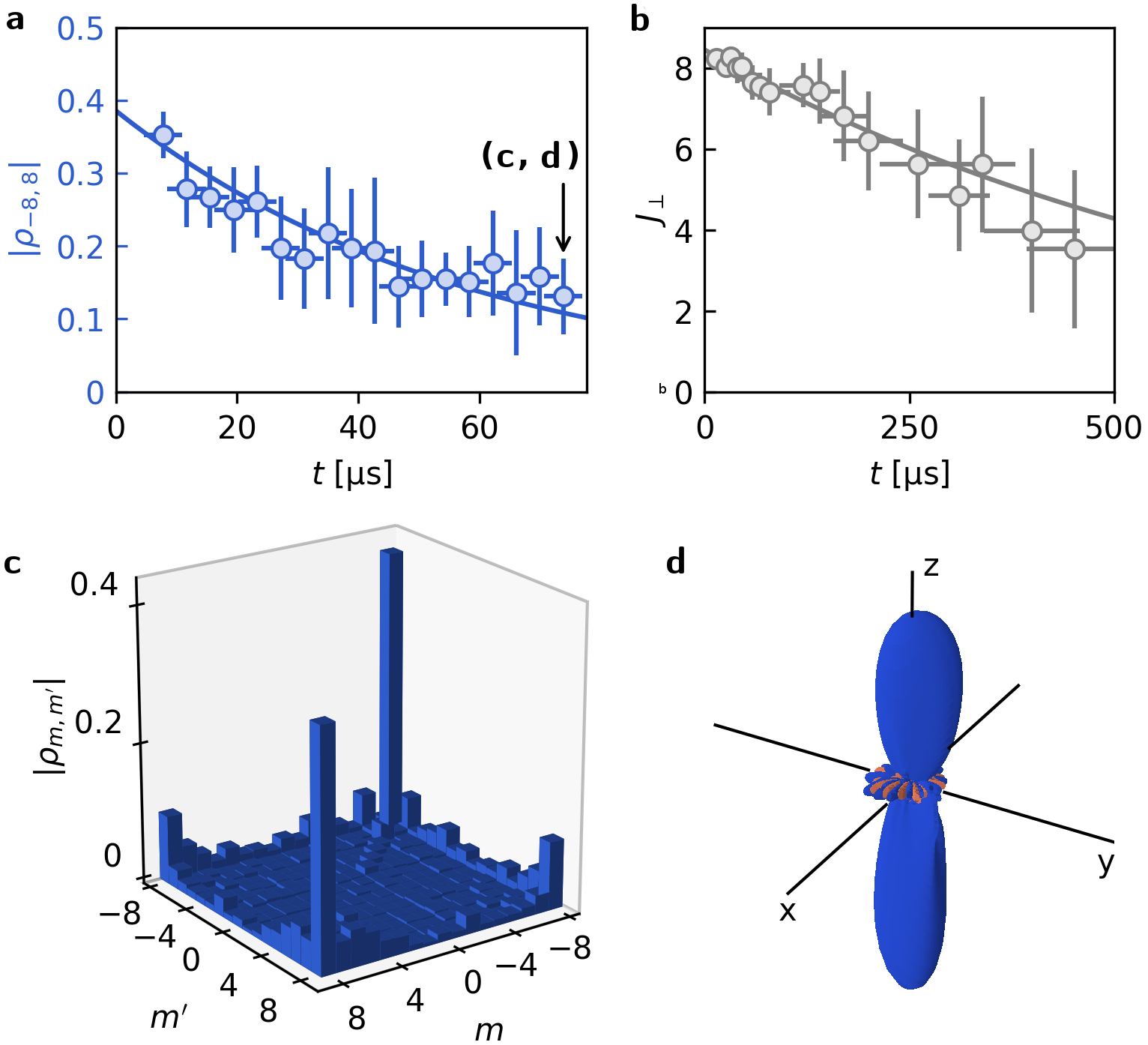}
\vspace{-5mm}
\caption{\textbf{Dephasing of coherences.}
(\textbf{a}) Evolution of the modulus of the extremal coherence $|\rho_{-8,8}|$ (blue circles)  calculated from the tomography of the superposition state after a wait time $t$. The horizontal error bars correspond to the standard deviation of the Larmor precession times required for tomography. Vertical error bars are the $1\sigma$ statistical error computed using a random-weight bootstrap method.
(\textbf{b}) Evolution of the mean transverse spin amplitude $J_\perp$ for an initial state $\ket{J}_x$  in the same magnetic field environment than for the data in a. The solid lines in a and b are exponential fits of the data. 
(\textbf{c} and \textbf{d}) Density matrix and Wigner function reconstructed for $t=\SI{70\pm3}{\micro\second}$, i.e. after a strong damping of coherences.
\label{fig_decoherence}}
\end{figure}

While we do not detect significant evolution of the populations $\Pi_m$, we observe a decrease of  the extremal coherence $|\rho_{-J,J}|$, of $1/\mathrm{e}$ decay time $\tau=\SI{58\pm4}{\micro\second}$, which we attribute to fluctuations of the ambient magnetic field. 
To calibrate such a dephasing process, we study the damping of the amplitude $J_\perp(t)$ of a coherent state, initially prepared in the state $\ket{J}_x$  and evolving under the applied magnetic field along $z$ and the ambient magnetic field fluctuations (see Methods). As shown in Fig.~\ref{fig_decoherence}b,  the transverse spin amplitude $J_\perp$ decays on a $1/\mathrm{e}$ timescale $\tau_0=\SI{740\pm80}{\micro\second}$, consistent with residual magnetic field fluctuations in the mG range.
The decoherence rate of the kitten state is thus enhanced by a factor  $\tau_0/\tau=13(2)$ compared to a coherent state, which illustrates the intrinsic fragility of mesoscopic coherent superpositions. 

Spin decoherence due to magnetic field fluctuations can be modeled similarly to the $T_2^*$ decay in nuclear magnetic resonance \cite{allen_optical_1975} 
(see the Supplementary Materials). Using a magnetic probe located close to the atom position, we measure shot-to-shot magnetic field fluctuations on a 0.5-mG range, but their variation on the $\sim\SI[number-unit-product=\text{-}]{100}{\micro\second}$ dephasing timescale remains negligible. In this regime, we expect the dephasing of the state $\cat$ to occur $2J=16$ times faster than for a coherent state, a value close to our measurement \cite{ferrini_noise_2010}. 

\footnotetext[8]{A purely Markovian evolution would be expected for white magnetic field noise, leading to phase diffusion behavior and a decoherence enhancement factor  $\tau_0/\tau=(2J)^2$ \cite{ferrini_noise_2010}, well above our measurement.}

Finally we plot in Fig.~\ref{fig_decoherence}c and d the reconstructed density matrix and its associated Wigner function for the wait time $t=\SI{70\pm3}{\micro\second}$. The weak amplitude of coherences and the shrinking of the negative regions in the Wigner function illustrate the dynamics towards an incoherent statistical mixture \cite{zurek_decoherence_2003}. 

\medskip
\noindent\textbf{Discussion}\\
\noindent 
In this work we use spin-dependent light shifts to  drive  the electronic spin $J=8$ of Dysprosium atoms under a non-linear one-axis twisting Hamiltonian. The observation of several collapses and revivals of quantum coherence shows that the spin dynamics remains coherent over a full period of the evolution. In particular,  the state produced after one quarter of the period consists of a coherent superposition between quasi-classical spin states  of opposite orientation, which can be viewed as a mesoscopic instance of  Schr\"odinger cat. While such coherent dynamics could be achieved with individual alkali atoms of  smaller spin size \cite{chaudhury_quantum_2007,fernholz_spin_2008,auccaise_spin_2015}, the realization of large-size coherent superpositions with ensembles of spin-1/2 particles is extremely challenging  \cite{monz_14-qubit_2011,yao_observation_2012}. The high fidelity of our protocol stems from the reduced size $2J+1$ of the available Hilbert space, that scales linearly with the effective distance $2J$ between the states involved in the superposition. Such scaling contrasts with the exponential scaling in the number of accessible states for ensembles of qubits, which dramatically increases the number of decoherence channels. Similarly, the full tomographic reconstruction of the produced quantum state also crucially relies on this limited size of the Hilbert space. Quantum state tomography of an equivalent 16-qubit ensemble remains inaccessible, unless restricting the Hilbert space  to the permutationally invariant subspace \cite{toth2010permutationally} or using compressed sensing for almost pure states \cite{gross_quantum_2010}.

We show that our kitten state provides a quantum enhancement of precision of 13.9(1.1), up to 87(2)\% of the Heisenberg limit. So far, such a high value could only be reached in ensembles of thousands of qubits based on multiparticle entanglement \cite{hamley_spin-nematic_2012,lucke_detecting_2014,bohnet_reduced_2014,hosten_measurement_2016,cox_deterministic_2016}. In such systems, while entanglement occurs between a large number of qubits, this number remains small compared  to the system size, far from the Heisenberg limit.  

Our method could be extended to systems of larger electronic spin $J$. Dysprosium being the optimum choice among all atomic elements in the electronic ground state, further improvement would require using highly excited electronic levels, such as Rydberg atomic states \cite{facon_sensitive_2016}, or using ultracold molecules \cite{frisch_ultracold_2015}.  
By increasing the atom density, one could also use interactions between $N$ atoms of spin $J$ to act on a collective spin of very large size $\mathcal{J}=NJ$, allowing to explore non-classical states of much larger size.

\medskip
\noindent\textbf{Methods}
\smallskip

{\small

\noindent\textbf{Sample preparation and detection}\\
\noindent
We use samples of about $9(1)\times10^4$ atoms of $^{164}$Dy, cooled to a temperature $T\simeq \SI{2}{\micro\kelvin}$ using laser cooling and subsequent evaporative cooling in an optical dipole trap \cite{dreon_optical_2017}.
The dipole trap has a wavelength $\lambda=1064\,$nm, resulting in negligible interaction with the atomic spin \cite{ravensbergen2018accurate}.
The samples are initially spin-polarized  in the absolute ground state $\ket{-J}_z$, with a bias field $B_z\simeq0.5$\,G along $z$, such that the induced Zeeman splitting largely exceeds the thermal energy. Before starting the light-induced spin dynamics, we  ramp the bias field down to the final value $B_z=18.5(3)$\,mG in 20\,ms. We checked that the promotion to higher spin states (with $m>-J$) due to dipole-dipole interactions remains negligible on this timescale. The optical trapping light is switched off right before the spin dynamics experiments. 

After the light-induced spin dynamics, we perform a Stern-Gerlach separation of the various spin components using  a transient magnetic field gradient  (typically 50\,G/cm during 2\,ms) with a large bias magnetic field along $z$. After a 3.5\,ms time of flight, the atomic density is structured as 17 separated profiles (see Fig.~\ref{fig_scheme}c), allowing to measure the individual spin projection probabilites $\Pi_m$ using resonant absorption imaging, where $m$ is the spin projection along $z$. The relative scattering cross-sections between $\ket{m}_z$ sub-levels are calibrated using samples of controlled spin composition.

Spin projection measurements along equatorial directions are based on spin rotations followed by a projective measurement along $z$. We apply a magnetic field pulse along $y$,  of temporal shape  $B_y(t)=B_{y}^{\mathrm{max}}\sin^2(\pi t/\tau)$, with $\tau=\SI{3}{\micro\second}$ and $B_{y}^{\mathrm{max}}$ adjusted to map the $z$ axis on the equator. Taking into account the static field along $z$, we expect the pulse to map the equatorial direction of azimutal angle $\phi_{\mathrm{i}}\simeq0.35\,$rad on the $z$ axis. An arbitrary angle $\phi = \phi_{\mathrm{i}}+\phi_{\mathrm{L}}$ can be reached using an additional wait time before the $B_y$ pulse, allowing for a Larmor precession of angle $\phi_{\mathrm{L}}$. The calculation of the angle $\phi_{\mathrm{L}}$uses the magnetic field component $B_z$  measured using an external probe, allowing to reduce the effect of shot-to-shot magntic field fluctuations. 

\medskip
\noindent\textbf{Spin dynamics modeling}\\
\noindent Quantitative understanding of the observed spin dynamics requires taking into account experimental imperfections. We include the linear Zeeman coupling induced by the magnetic field applied along $z$  (see Eq.~\ref{eq_H}), leading to a small Larmor rotation on the typical timescales used for the light-induced spin dynamics. We also account for a small angle mismatch between the quantization field and the $z$ axis and the inhomogeneity of the atom-light coupling due to the finite extent of the atomic sample (see the Supplementary Materials).

\medskip
\noindent\textbf{Quantum state tomography}
\\
\noindent The density matrix of the kitten state is determined from a least-square fit of the measured spin projection probabilities $\Pi_m$ along $z$ and $\Pi_m(\phi)$ on equatorial directions \cite{klose_measuring_2001}. We uniformly sample the equatorial plane using a set of azimutal angles $\phi\in[\phi_0,\phi_0+\pi]$. The procedure thus requires variable spin rotation durations (on average $\simeq\SI{10}{\micro\second}$), which limits the quality of the tomography due to dephasing. To reduce its effect, we use the magnetic field values  measured for each experimental with an external probe to compensate for part of the dephasing, which increases the quality of the tomography and extents the coherence times by a factor $\simeq3$.
The robustness of the method with respect to measurement noise and finite sampling is tested using a random-weight bootstrap method, from which we define  the statistical error bars in Fig.~\ref{fig_decoherence}.

\medskip
\noindent\textbf{Calibration of dephasing}
\\
\noindent To calibrate the dephasing of coherences due to magnetic field fluctuations, we perform a Ramsey experiment using coherent spin states. We start in the ground state $\ket{-J}_z$, that we bring on the equator using a $\pi/2$ magnetic field pulse applied along $y$. We then let the spin precess around $z$ for a duration $t$, and subsequently perform a second $\pi/2$ pulse before performing a spin projection measurement along $z$. We observe Ramsey oscillations of the magnetization $m_z(t)=J_\perp(t)\cos(\omega_{\mathrm{L}} t+\phi)$, where the local oscillation contrast $J_\perp(t)$ corresponds to the transverse spin amplitude shown in Fig.~\ref{fig_decoherence}b. 

}

%\bibliographystyle{Science}
%\bibliography{references}

\medskip
\filbreak
\noindent\textbf{Acknowledgments}\\
\noindent This work is supported by PSL University (MAFAG project) and European Union (ERC UQUAM \& TOPODY, Marie Curie project 661433).  We thank F. Gerbier, R. Lopes and P. Zoller for fruitful discussions.

\medskip
\noindent\textbf{Author contributions}\\
\noindent 
T.C., L.S., C.B., A.E., V.M. and D.D. carried out the experiment. J.D and S.N.
supervised the project. All authors contributed to the discussion, analysis of the results
and the writing of the manuscript.

\clearpage
\noindent\textbf{SUPPLEMENTARY MATERIALS}

\renewcommand{\thefigure}{S\arabic{figure}}
\renewcommand{\thetable}{S\arabic{table}}
\renewcommand{\theequation}{S\arabic{equation}}
\setcounter{figure}{0}
\setcounter{table}{0}
\setcounter{equation}{0}

\medskip
\noindent\textbf{Ideal spin dynamics}\\
We present here the expected dynamics for a pure $\hat J_x^2$ coupling. By decomposing the initial state $\ket{-J}_z$ on the $x$ basis $\{\ket{m}_x\}$, we find the evolved state as
\begin{equation}\label{eq_ideal_dynamics}
\ket{\psi(t)}=\frac{1}{2^J}\sum_m\E^{-\I m^2\omega t}(-1)^m\sqrt{{{2J}\choose{J+m}}}\ket{m}_x.
\end{equation}
The magnetization and spin projection variance along $z$ were calculated analytically in Ref.~\cite{kitagawa_squeezed_1993}, as
\begin{align}
m_z(t)=&-J\, \left[\cos\left(\omega t\right)\right]^{2J-1},\label{eq_mz_exact}\\
\Delta J_z^2(t)= &\; J^2\left(1-\left[\cos\left(\omega t\right)\right]^{2(2J-1)}\right)\nonumber\\
&-\frac{J(J-\frac{1}{2})}{2}\left(1-[\cos(2\omega t)]^{2J-2}\right),\label{eq_varz_exact}
\end{align}
corresponding to the dashed red lines in Fig.~\ref{fig_evolution}b and c.

\begin{figure}
\hspace*{-0.0\linewidth}\includegraphics[width=1.0\linewidth]{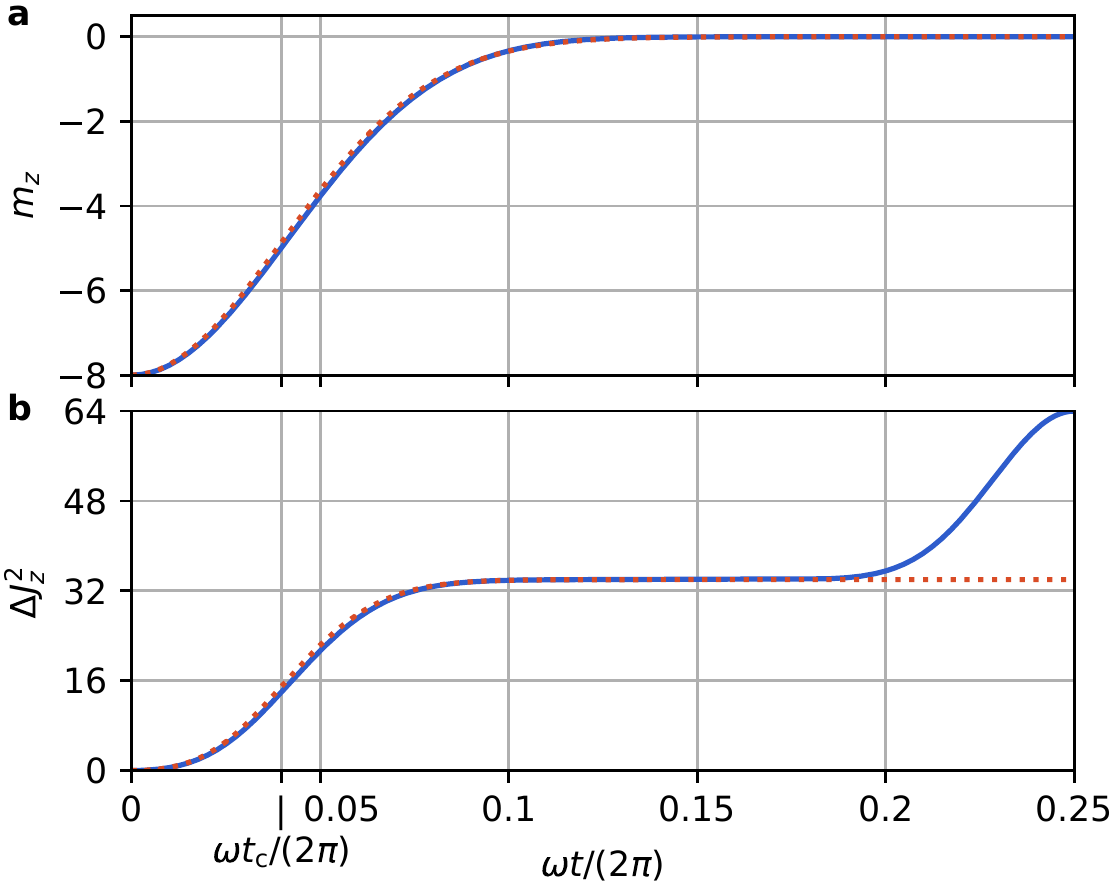}
\vspace{-4mm}
\caption{\textbf{Evolution of the first spin moments during the collapse.}
(\textbf{a}, \textbf{b}) Evolution of the magnetization $m_z$ (\textbf{a}) and of the spin projection variance $\Delta J_z^2$ (\textbf{b})  under a pure $\hat J_x^2$ coupling. The solid blue lines corresponds to the exact expressions (\ref{eq_mz_exact}) and (\ref{eq_varz_exact}). The red dotted lines represent their gaussian approximations given by Eqs.~(\ref{eq_mz_G}) and (\ref{eq_varz_G}).
\label{fig_evolution_S}}
\end{figure}

The  collapse of quantum coherence results from the relative dephasing between the various $m^2\omega t$ phases. For large $J$ values, the evolution of $m_z$ and $\Delta J_z^2$ are well captured by a gaussian decay on a timescale $t_c=1/(\sqrt{2J}\omega)$, as
\begin{align}
m_z^{\mathrm{G}}(t)\simeq&-J\exp[-(t/t_c)^2/2],  \label{eq_mz_G}\\
{[\Delta J_z^{\mathrm{G}}(t)]^2}\simeq& \frac{J(J+\frac{1}{2})}{2}-[m_z^{\mathrm{G}}(t)]^2\nonumber\\
&\!\!\!\!\!\!\!\!\!\!\!\!\!\!\!\!\!+\frac{J(J+\frac{1}{2})}{2}\exp\left[-\frac{2J^2-2J+1}{J(J-\frac{1}{2})}(t/t_c)^2\right].\label{eq_varz_G}
\end{align}
We show in Fig.~\ref{fig_evolution_S} the evolution of the magnetization and spin projection variance for $J=8$. We find that the gaussian approximations of Eqs.~(\ref{eq_mz_G}) and (\ref{eq_varz_G}) reproduce very well the exact formulas (\ref{eq_mz_exact}) and (\ref{eq_varz_exact}) during  the entire collapse ($0<\omega t<0.3\pi$).

This  gaussian decay of the magnetization and spin projection variance ceases to be valid close to $\omega t=\pi/2$, at which all even (odd) $m$ phase factors get rephased together, leading to a revival of coherence. Similar revivals of coherence are expected to occur for $\omega t=n\pi/2$ ($n$ integer), corresponding to quantum states
\begin{align}
\ket{\psi}&=\frac{1}{\sqrt{2}}\left(\E^{\I n\pi/4}\ket{-J} + \E^{-\I n\pi/4} \ket{J})\right), \quad n\;\mathrm{odd} \\
\ket{\psi}&=\ket{(-1)^{n/2}J}, \quad n\;\mathrm{even} .
\end{align}

\medskip
\noindent\textbf{Spin dynamics modeling}
\\
\noindent The non-linear spin dynamics results from the spin-dependent light shifts caused by laser light whose wavelength is close to the  optical transition at \mbox{$\lambda_0=626$\,nm}, that couples the electronic ground state to an excited level of angular momentum \mbox{$J'=J+1$}. The light frequency is detuned from resonance by \mbox{$\Delta=-2\pi\times 1.5\,$GHz}, a value chosen to maximize the light shift amplitude while keeping incoherent light scattering negligible over the timescale of our experiments. For such a detuning the contribution from other optical resonances to the light shift is negligible.

The light-induced spin coupling, whose structure depends on the light polarization $\unitu$, is obtained using second-order perturbation theory as
\begin{align*}
\hat V=&V_0\Bigg\{
\frac{2J+3}{3(2J+1)}\hat{\mathds{1}}
-i\frac{2J+3}{2(J+1)(2J+1)}[\unitu^*\times\unitu]\cdot\hat\Jbold\\
&-\frac{3[(\unitu^*\cdot\hat\Jbold)(\unitu\cdot\hat\Jbold)+(\unitu\cdot\hat\Jbold)(\unitu^*\cdot\hat\Jbold)]-2\hat\Jbold^2}{6(J+1)(2J+1)}\Bigg\}
,\\
V_0=&\frac{3\pi c^2\Gamma}{2\omega_0^3\Delta}I,
\end{align*}
where $c$ is the speed of light, $\Gamma=\SI{0.85\pm0.03}{\micro\second^{-1}}$ is the resonance linewidth \cite{gustavsson_lifetime_1979}, $\omega_0=2\pi c/\lambda_0$, and $I$ is the light intensity. The $\hat J_x^2$ coupling used to create the  superposition state  is achieved for a linear polarization $\unitu=\unitx$. We discuss below several types of experimental imperfections affecting the spin dynamics.

\smallskip
\noindent \textit{Finite extent of the atomic sample}\\
\noindent We first take into account the variation of intensity and polarization  over the atomic sample. We focus  on the atomic sample a  gaussian beam linearly polarized along $x$ and propagating along $y$, of  waist $w\simeq\SI{50}{\micro\meter}$ at the atom position. Due to the beam focusing, we expect a slight polarization ellipticity away from the optical axis $\unitu\simeq\unitx+\I \theta x/w\unity$, where $\theta=\lambda/(\pi w)\simeq4\,$mrad is the beam divergence.  The dynamics shown in Fig.~\ref{fig_evolution} is consistent with a $1/\mathrm{e}$ cloud size $\sigma=\SI{7.3\pm0.3}{\micro\meter}$, in agreement with the size calculated from the trap geometry and the gas temperature. Given this cloud extent, we expect r.m.s. intensity variations between atoms $\delta I/I=6\%$ and an ellipticity typically corresponding to a Stokes parameter $s_3=10^{-3}$. 

\smallskip
\noindent \textit{Quantization magnetic field}\\
\noindent We  include the effect of the applied magnetic field of amplitude $B=18.5(3)\,$mG, leading to a linear Zeeman coupling. We fit its orientation from the measured spin dynamics, consistent with an angular mismatch between the quantization field direction $\unitb$ (of components $[0.09,-0.11,0.98]_{x,y,z}$) and the $z$ axis.

\smallskip
\noindent \textit{Imperfect spin polarization}\\
\noindent The atomic gas is prepared in the absolute ground state $\ket{-J}_z$ under a strong magnetic field $B_z=0.5\,$G. The field is ramped to the final value $B_z=18.5(3)$\,mG in 20\,ms, during which dipole-dipole interactions lead to a slight promotion of $\simeq3\%$ of the atoms into the state $\ket{-J+1}_z$. We take into account the spin dynamics undergone by these atoms. 

\smallskip
\noindent \textit{Light shift correction to second-order perturbation theory}\\
\noindent Given the small detuning from resonance, we calculate the first correction to second-order perturbation theory in the light shift. For a light field linearly polarized along $x$, we obtain the expression
\[
\hat V=\hbar\omega\!\left\{\hat J_x^2+\!\frac{\omega}{\Delta}\left[(2J^2\!+3J+1)\hat J_x^2+\hat J_x^4\right]+\mathcal{O}[(\omega/\Delta)^2]\right\}\!\!.
\]
The first correcting term leads to a renormalization of the coupling frequency $\omega$, corresponding to a $\sim20\%$ reduction of $\omega$ for our experimental parameters. The expected (renormalized) value $\omega=2\pi\times1.95(10)\,$MHz  agrees well with the value $\omega=2\pi\times1.98(1)\,$MHz fitted from the spin dynamics. The second term $\propto \hat J_x^4$ leads to a slight modification of the spin dynamics, but its effect remains below the experimental noise.

\smallskip
\noindent \textit{Light intensity response time}\\
\noindent The light pulse shape is controlled using an acousto-optic modulator, leading to a finite response time in the 10-100\,ns range. By solving numerically the spin dynamics with the actual pulse shape, we estimate that the finite response time leads to minor differences compared to square pulses of same area. 

\smallskip
\noindent \textit{Incoherent light scattering}\\
\noindent We model the effect of incoherent light scattering, taking into account Rayleigh and Raman scatterings using a Monte Carlo wavefunction method \cite{dalibard_wave-function_1992}. For the detuning chosen in our experiments, we estimate a light scattering probability of 0.7\% for the light pulse duration required to produce the superposition state.

We compare the effects of the different imperfections in Tab.~\ref{tab}. For each imperfection, we calculate the resulting decrease of the metrological gain $G$ with respect to the maximum value of $2J=16$. The dominant imperfection stems from the finite extent of the atomic gas, leading to inhomogeneous light intensity and to polarization ellipticity. Combining all effects together, we estimate a maximum metrological gain of $G=14.5$, consistent with the measured value $G=13.9(1.1)$. 

\begin{table}[t]
\begin{tabular}{lc}
\hline
\hline
Imperfection & correction to $G$\\
\hline
Intensity inhomogeneity &-1.43\\
Polarization ellipticity &-0.41\\
$B_z$ field amplitude&-0.22\\
$\unitb$ and $\unitz$ angle mismatch&-0.34\\
Imperfect initial state polarization &-0.06\\
$\hat J_x^4$ correction & -0.18\\
Light intensity response time & -0.18\\
Incoherent light scattering & -0.09\\
\hline
Combined correction& -1.52\\
\hline
\hline
\end{tabular}
\caption{\label{tab}Balance of experimental imperfections in the reduction of the metrological gain.}
\end{table}

\medskip
\noindent\textbf{Metrological gain versus measurement scheme}
\\
\noindent We discuss in the main text two methods to probe magnetic fields using the kitten state, based on the evolution of the projection probability distributions $\Pi_m(\phi)$ along equatorial directions shown in Fig.\;\ref{fig_Ramsey}a. The first method uses the oscillatory behavior of the parity $P(\phi)$ of these distributions,  and the second is based on the variation of the Hellinger distance $d_{\mathrm{H}}(\phi,\phi')$ between probability distributions for different phases $\phi$ and $\phi'$.

A similar analysis can be performed using the non-linear detection scheme, which leads to the probability distributions $\Pi_m(\phi)$ shown in Fig.\;\ref{fig_Ramsey}b. We first exploit the measured magnetization oscillations shown in Fig.~\ref{fig_Ramsey}c. We evaluate the measurement precision using the general formula (\ref{eq_Delta_phi}) with the observable \mbox{$\hat{\mathcal{O}}=\hat J_z$}, leading to the expression of the metrological gain 
\mbox{$
G=2J(A/\Delta J_z)^2,
$}
where $A=6.0(2)$ is the oscillation amplitude and $\Delta J_z^2=49(1)$ is the spin projection variance (measured for the $m_z\simeq0$ data). A second measurement scheme consists in quantifying the variations of the  probability distributions $\Pi_m(\phi)$ using the Hellinger distance, following the same procedure than for the data of Fig.~\ref{fig_Ramsey}a.

We show in Tab.~\ref{tab_G} the metrological values obtained from the four measurement schemes discussed above. For both data sets we find that exploiting the variations of the Hellinger distance $d_{\mathrm{H}}(\phi,\phi')$ leads to the highest $G$ values, both of them being compatible with the upper bound \mbox{$2\Delta J_z^2/J=14.3(1)$}.

\begin{table}[t]
\begin{tabular}{lc}
\hline
\hline
Data of Fig.~\ref{fig_Ramsey}a & $G$\\
\hline
Parity oscillation &8.8(4)\\
Variations of probability distributions &13.9(1.1)\\
\hline
\hline
Data of Fig.~\ref{fig_Ramsey}a & $G$\\
\hline
Magnetization oscillation &11.2(1.3)\\
Variations of probability distributions &14.0(9)\\
\hline
\hline
\end{tabular}
\caption{\label{tab_G}Metrological gain values corresponding to the four measurement schemes discussed in the main text.}
\end{table}

\medskip
\noindent\textbf{Dephasing due to magnetic field fluctuations}
\\
\noindent As discussed in the main text, we mainly attribute the observed decoherence  to magnetic field fluctuations along $z$. The dephasing due to this classical noise can be modeled using standard techniques from nuclear magnetic resonance \cite{allen_optical_1975,luczka_spin_1990,palma_quantum_1996}. We write the magnetic field as
$
B_z(t)=B_z^{0}+b(t)
$
with $\langle b(t) \rangle=0$. Such a field induces a noise in the Larmor rotation angle
\[
\delta\phi(t)=\frac{\muB g_J}{\hbar}\int_0^t\mathrm{d}t' b(t').
\]
For a coherent state prepared on the equator, this noise leads to the decay of transverse spin $J_\perp(t)=J\left<\E^{\I\delta\phi(t)}\right>$. For a superposition state $\cat$ it reduces the extremal coherence as $|\rho_{-J,J}|=\left<\E^{\I2J\delta\phi(t)}\right>$.

A purely Markovian evolution would be expected for white magnetic field noise, leading to phase diffusion behavior \cite{ferrini_noise_2010}.
The Markovian approximation corresponds to phase diffusion without memory, corresponding to \mbox{$\langle b(t)b(t')\rangle=2D\,\delta(t-t')$} \cite{van_kampen_stochastic_1992}. For gaussian noise statistics, we use  \mbox{$\left<\E^{\I n\delta\phi(t)}\right>=\E^{-n^2\left<\delta\phi(t)\right>^2/2}$}, with a diffusive phase noise  \mbox{$\left<\delta\phi(t)\right>^2=2D(\muB g_J/\hbar)^2t$}, leading to exponential damping of coherences, of $1/\E$ times  \mbox{$\tau_0=(\hbar/\muB g_J)^2D$} for the transverse spin $J_\perp$ and $\tau=\tau_0/(2J)^2$ for the extremal coherence  $|\rho_{-J,J}|$.

In our experiment, we rather expect typical magnetic field variations to occur on timescales much larger than the $\sim\SI{100}{\micro\second}$ decoherence timescale. In this regime, the magnetic field can be considered as static during a single realization of the experiment, and decoherence arises from shot-to-shot fluctuations. We then expect $\delta\phi(t)=\muB g_Jb t/\hbar$, which obviously leads to the relationship 
\[
\left<\E^{\I2J\delta\phi(t)}\right>=\left<\E^{\I\delta\phi(2Jt)}\right>.
\] 
This relationship implies that the damping of extremal quantum coherences after a duration $t$ is equal to the damping of the coherence of coherent states after a duration $2J t$, consistently with our observations.

Since each atom carries a magnetic moment $\sim 10\,\muB$, we also expect an additional magnetic field created by the atomic sample itself, but we estimate its contribution to the coherence damping rate to be one order of magnitude smaller than external magnetic field effects.

\end{document}